\documentclass[
twocolumn,
print,
showpacs,
preprintnumbers,
nofootinbib,
 amsmath,amssymb,
 prd,
]{revtex4-1}
\usepackage{amssymb}
\usepackage{graphicx}
\usepackage{dcolumn,color}
\usepackage{bm}
\usepackage{epstopdf}
\usepackage{epsfig}
\usepackage{color,framed}
\usepackage{amsmath,amsthm,amssymb}

\begin{document}

\title{Cosmological twinlike models with multi scalar fields}

\author{Yuan Zhong$^{1}$, Chun-E Fu$^{1}$, Yu-Xiao Liu$^{2}$\footnote{The corresponding author:~liuyx@lzu.edu.cn}}
\affiliation{$^1$School of Science, Xi'an Jiaotong University,
Xi'an 710049, People's Republic of China}
\affiliation{$^2$Institute of Theoretical Physics, Lanzhou University,
           Lanzhou 730000, People's Republic of China}

\begin{abstract}
We consider cosmological models driven by several canonical or noncanonical scalar fields. We show how the superpotential method enables one to construct twinlike models for a particular canonical model from some noncanonical ones. We conclude that it is possible to construct twinlike models for multi-field cosmological models, even when the spatial curvature is nonzero. This work extends the discussions of [D. Bazeia and J. D. Dantas, Phys. Rev. D, 85 (2012) 067303] to cases with multi scalar fields and with non-vanished spatial curvature, by using a different superpotential method.
\end{abstract}

\maketitle

\section{Introduction}
Scalar field plays an important role in cosmology. It can be applied either to trigger the inflation that happened in the early Universe~\cite{LiddleLyth2000}, or to be a dark energy candidate such as quintessence, to explain the recently observed accelerated expansion of our Universe (see for example, Refs.~\cite{RatraPeebles1988,CaldwellDaveSteinhardt1998,FriemanTurnerHuterer2008}). In quintessence inflation model it is even possible to use only one scalar to describe both the early inflation and the dark energy~\cite{DimopoulosValle2002}. Canonical scalar field is an economic choice both for constructing inflation and dark energy models. With the development of effective field theory, however, people pay more attention to nonrenormalizable field models. The K-field is a typical class of noncanonical scalar field models, whose Lagrangian $\mathcal{L}(\phi,X)$ is an arbitrary function of the scalar $\phi$ and its kinetic term $X=-\frac12g_{\mu\nu}\nabla^\mu\phi\nabla^\nu\phi$.
The K-field was initially introduced to describe early inflation~\cite{Armendariz-PiconDamourMukhanov1999,GarrigaMukhanov1999,Armendariz-PiconMukhanovSteinhardt2001}, and was later been applied in the study of topological defects~\cite{BazeiaLosanoMenezesOliveira2007,BazeiaLosanoMenezes2008,ZhongLiu2014}, brane world~\cite{BazeiaGomesLosanoMenezes2009,BazeiaLobLosanoMenezes2013,ZhongLiu2013,ZhongLiuZhao2014a,BazeiaLobaoMenezes2015}, et al.

In 2010, a new interesting application of K-field was reported in Ref.~\cite{AndrewsLewandowskiTroddenWesley2010}. The authors found that under some conditions a K-field model can have the same background topological defect solutions than a canonical model. Such a K-field model is dubbed as a ``doppelg\"anger", or a twinlike model of the canonical one. Despite of the equivalence at the background level, twinlike models usually have different linear perturbation structures, and are distinguishable in principle (see also Refs.~\cite{AdamQueiruga2011,BazeiaDantasGomesLosanoMenezes2011}). However, it was later found that when some further conditions are satisfied, the twinlike models can even have the same linear structure~\cite{BazeiaMenezes2011}. Such a pair of twinlike models is referred to as special twinlike models~\cite{AdamQueiruga2012,ZhongLiu2015}.

It is interesting to construct twinlike models in varies kinds of scalar field theories. So far, twinlike models have been constructed for self-dual Abelian-Higgs model~\cite{BazeiaHoraMenezes2012}, for compacton solutions~\cite{BazeiaLobMenezes2012} and for thick brane world solutions~\cite{BazeiaLobMenezes2012,BazeiaDantasGomesLosanoMenezes2011,ZhongLiu2015}.
In addition to models with only one scalar, there are also works on multi-field twinlike models~\cite{BazeiaLobaoLosanoMenezes2014}.

Since scalar field plays an important role in cosmology, it is also interesting to consider cosmological twinlike models. The first cosmological twinlike model was constructed in Ref~\cite{BazeiaDantas2012}. Using the first-order formalism\footnote{Also known as the superpotential method. See for example Refs.~\cite{BazeiaLosanoRodrigues2006,BazeiaGomesLosanoMenezes2006} for the application of this method in the study of dark energy.}, the authors successfully constructed cosmological twinlike models in the case with a single scalar field and with vanished spatial curvature. Due to the form of the first-order formalism, the authors of Ref.~\cite{BazeiaDantas2012} failed to find twinlike models for cosmological models with nonzero spatial curvature. They concluded that the appearance of the spatial curvature forbids the construction of twinlike models in the cosmological scenario.

In this paper, however, we will show that it is indeed possible to construct cosmological twinlike models no matter the spatial curvature is vanished or not. To achieve this goal, we will apply a new first-order formalism, which is more convenient for constructing twinlike models as compared to the one used in Ref.~\cite{BazeiaDantas2012}. We will also apply our first-order formalism to the case with multi scalar fields. Our study indicates that it is possible to construct twinlike models in multi-field models, even in a curved space-time. This study compensates the work of Ref.~\cite{BazeiaLobaoLosanoMenezes2014}, where multi-field twinlike models were constructed only in two-dimensional flat space-time.

This paper is organized as follows. In Sec.~\ref{sectionCosTwinTwo} we first introduce the model and our conventions. In Sec.~\ref{sectionCosTwinThree}, we establish the first-order formalism for cosmological models with $n$ canonical scalar fields and with an arbitrary spatial curvature. Three types of twinlike models are constructed for these canonical models in Sec.~\ref{sectionCosTwinFour} by using the first-order formalism given in Sec.~\ref{sectionCosTwinThree}. Some explicit examples of twinlike models are given in Sec.~\ref{sectionCosTwinFive}. Our results will be summarized in Sec.~\ref{sectionCosTwinSix}.

\section{The model}
\label{sectionCosTwinTwo}

In cosmological models, the geometry of the space-time is described by the Friedmann-Robertson-Walker (FRW) metric
\begin{eqnarray}
&&ds^2=g_{\mu\nu}dx^\mu dx^\nu\nonumber\\
&=&-dt^2+a^2(t)\left[\frac{dr^2}{1-k r^2}+r^2(d\theta^2+\sin^2\theta d\varphi^2)\right],
\end{eqnarray}
where $a(t)$ is the scale factor, and the constant $k=1,$ $0$, or $-1$ corresponds to spherical, flat, or hyperbolic geometry, respectively. Space-time indices are denoted by Greek letters $\mu, \nu=0,1,2,3$.

The action of our model is composed by two parts:
\begin{eqnarray}
\label{actionGr}
S=S_{HE}+S_M.
\end{eqnarray}
As usual, the gravitational part is given by the Hilbert-Einstein action:
\begin{eqnarray}
S_{HE}=\frac{1}{16\pi G}\int d^4 x \sqrt{-g}R,
\end{eqnarray}
where $G$ is the gravitational coupling and $g=\det g_{\mu\nu}$ is the determinant of the metric. For convenience, and to compare with Ref.~\cite{BazeiaDantas2012}, we take $4 \pi G=1$.
The matter part is described by the following action:
\begin{eqnarray}
\label{eqSm}
S_{M}=\int d^4 x \sqrt{-g}\mathcal{L}(\mathcal{G}_{IJ},X^{IJ},\phi^I),
\end{eqnarray}
where $\mathcal{G}_{IJ}=\mathcal{G}_{IJ}(\phi^K)$ is the metric of the field space such that $\phi_I=\mathcal{G}_{IJ}\phi^J$, and $X^{IJ}=-g^{\mu\nu}\partial_\mu\phi^I\partial_\nu\phi^J/2$ is the kinetic term for $n$ scalar fields $\phi^I=\phi^I(t)$ with $I,J,K=1,2,\cdots,n$.

From the Hamiltonian variation principle $\delta S/\delta g^{\mu\nu}=0$, one can easily obtain the Einstein equations:
\begin{eqnarray}
\label{EqRR2T}
R_{\mu\nu}- \frac{1}{2}g_{\mu\nu}R =2 {T_{\mu\nu}},
\end{eqnarray}
where the energy-momentum tensor is defined as
\begin{eqnarray}
T_{\mu\nu}&=&-2\frac1{\sqrt{-g}}\frac{\delta S_M}{\delta g^{\mu\nu}}.
\end{eqnarray}
In particular, as Eq.~\eqref{eqSm} is considered, we have
\begin{eqnarray}
T_{\mu\nu}=\mathcal{L}_{X^{IJ}}\partial_\mu\phi^I\partial_\nu\phi^J+g_{\mu\nu}\mathcal{L}.
\end{eqnarray}
Here, we have defined $\mathcal{L}_{X^{IJ}}=\partial \mathcal{L}/\partial X^{IJ}$.

After a simplification, the Einstein equations give
\begin{eqnarray}
{H^2} &=&\frac{2}{3}\left({{\cal L}_{{X^{IJ}}}}{{\dot \phi }^I}{{\dot \phi }^J} - {\cal L}\right) - \frac{k}{{{a^2}}},\\
\dot H &=& \frac{k}{{{a^2}}} - {{\cal L}_{{X^{IJ}}}}{{\dot \phi }^I}{{\dot \phi }^J}.
\end{eqnarray}
Note that an over dot is used to represent the derivative with respect to $t$, and $H=\dot{a}/a$ is the Hubble parameter. The Einstein equations can also be rewritten in terms of the energy density $\rho$ and the pressure $p$ of the matter fields:
\begin{eqnarray}
\label{EqH}
{H^2} &=&\frac{2}{3}\rho - \frac{k}{{{a^2}}},\\
\label{EqdotH}
\dot H &=& \frac{k}{{{a^2}}} -(\rho+p),
\end{eqnarray}
where $\rho$ and $p$ are defined as
\begin{eqnarray}
\label{EqrhoL}
 \rho &=&-T_0^0={{\cal L}_{{X^{IJ}}}}{{\dot \phi }^I}{{\dot \phi }^J} - {\cal L},\\
 \label{EqpL}
p &=&T_1^1={\cal L}.
\end{eqnarray}
Another important quantity in cosmology is the deceleration parameter
\begin{eqnarray}
q\equiv-( 1+\frac{\dot{H}}{H^2}).
\end{eqnarray}

\section{The canonical model and the first-order formalism}
\label{sectionCosTwinThree}
In this section, we consider the canonical model, for which the field space metric $\mathcal{G}_{IJ}=\delta_{IJ}$, and the Lagrangian density of the scalar fields takes the form
\begin{eqnarray}
\label{EqbarL}
 &&\bar{\mathcal{L}}=\delta_{IJ}\dot{\phi}^I\dot{\phi}^J/2-V(\{\phi^I\}), \\
 \label{EqbarLX}
 &&\bar{\mathcal{L}}_{X^{IJ}}=\delta_{IJ}.
\end{eqnarray}
Here an over bar denotes the quantities of the canonical model.
In this case,
\begin{eqnarray}
\bar{\rho}&=& \frac{1}{2}{\delta _{IJ}}{{\dot \phi }^I}{{\dot \phi }^J} + V,\\
\bar{p}&=& \frac{1}{2}{\delta _{IJ}}{{\dot \phi }^I}{{\dot \phi }^J} - V,
\end{eqnarray}
and the Einstein equations reduce to
\begin{eqnarray}
\label{SMeq1}
{H^2} &=& \frac{1}{3}{\delta _{IJ}}{{\dot \phi }^I}{{\dot \phi }^J} + \frac{2}{3}V - \frac{k}{{{a^2}}},\\
\label{SMeq2}
\dot H &=& \frac{k}{{{a^2}}} - {\delta _{IJ}}{{\dot \phi }^I}{{\dot \phi }^J}.
\end{eqnarray}

To solve these equations, we introduce the following first-order formalism\footnote{This is a multi-field generalization of the first-order formalism in Ref.~\cite{BazeiaLosanoRodrigues2006}, where the authors first established the same first-order formalism for cosmological models with a single scalar.}:
\begin{eqnarray}
\label{phiW}
\dot{\phi}^I&=&\delta^{IJ}\frac{\partial W}{\partial \phi^J},\\
\label{HW}
H&=&-W+\alpha k Z,
\end{eqnarray}
where $\alpha>0$ is a positive parameter, $W$ and $Z$ are functions of $\{\phi^I\}$, and are called as the superpotentials. For $n=1$, it is quite easy to discern the difference between the above first-order formalism from those in Ref.~\cite{BazeiaDantas2012} (see also Refs.~\cite{BazeiaLosanoRodrigues2006,BazeiaGomesLosanoMenezes2006}), where
\begin{eqnarray}\label{EqBazeiaPhi}
\dot{\phi}&=&\alpha k Z-\frac{\partial W}{\partial \phi},\\
\label{EqBazeiaW}
H&=&W.
\end{eqnarray}
As analyzed in Ref.~\cite{BazeiaDantas2012}, such a first-order formalism leads difficulties for constructing twinlike models with $k\neq 0$. Besides, it is also not easy to generalize Eqs.~\eqref{EqBazeiaPhi} and \eqref{EqBazeiaW} to multi-field models.

Using the first-order formalism \eqref{phiW} and \eqref{HW} and the Einstein equations \eqref{SMeq1} and \eqref{SMeq2}, one can express $V,~\rho,~p$, and $q$ in terms of $W$ and $Z$:
\begin{eqnarray}
\label{VW}
V&=&\frac{3}{2} (W-\alpha k  Z )^2+\frac{\delta^{IJ}}{2}\frac{\partial W}{\partial\phi^I}\left(3 \alpha k \frac{\partial Z}{\partial\phi^J}-\frac{\partial W}{\partial\phi^J}\right),\nonumber\\
\bar{\rho} &=& \frac{3}{2}\left[{(W - \alpha kZ)^2} + \alpha k{\delta ^{IJ}}\frac{{\partial W}}{{\partial {\phi ^I}}}\frac{{\partial Z}}{{\partial {\phi ^J}}}\right],\nonumber\\
\bar{p} &=&  - \frac{3}{2}{(W - \alpha kZ)^2}-{\delta ^{IJ}}\frac{{\partial W}}{{\partial {\phi ^I}}}\left( {\frac{3\alpha k}{2}\frac{{\partial Z}}{{\partial {\phi ^J}}} - \frac{{\partial W}}{{\partial {\phi ^J}}}} \right),\nonumber\\
\bar{q}&=&\frac{\delta^{IJ}}{(W-\alpha k Z)^2}\frac{\partial (W-k\alpha Z)}{\partial {\phi ^I}}\frac{{\partial W}}{{\partial {\phi ^J}}}-1.
\end{eqnarray}

Note that the superpotentials $W$ and $Z$ are not independent. To see this, let us substitute the first-order equations \eqref{phiW} and \eqref{HW} into Eq.~\eqref{SMeq2}, after a simplification we get
\begin{equation}
\label{eqaWZ}
a^{-2}=\alpha \delta^{IJ}\frac{\partial Z}{\partial \phi^I}\frac{\partial W}{\partial \phi^J}.
\end{equation}
Taking the derivative of Eq.~\eqref{eqaWZ} with respect to $t$, and using Eqs.~\eqref{phiW} and \eqref{HW}, one immediately obtains the following constraint for the superpotentials:
\begin{eqnarray}
\label{Eqconstraint}
 &&0= \delta^{IJ} \left\{2(W-\alpha k Z)\frac{\partial Z}{\partial\phi^I}\frac{\partial W}{\partial\phi^J}\right.\nonumber\\
 && \left.-\delta^{KL}\frac{\partial W}{\partial\phi^L}
 \left(\frac{\partial W}{\partial\phi^J}\frac{\partial^2 Z}{\partial\phi^I\partial\phi^K}
 +\frac{\partial Z}{\partial\phi^I}\frac{\partial ^2 W}{\partial\phi^J\partial\phi^K}\right)\right\}.
\end{eqnarray}

The first-order formalism makes it easier to find analytical solutions for cosmological models. Some examples can be found in Refs.~\cite{BazeiaGomesLosanoMenezes2006,BazeiaLosanoRodrigues2006} for the dark energy. Our aim for this letter, however, is to construct twinlike models for the canonical model $\bar{\mathcal{L}}$. That is to find models whose Lagrangian $\mathcal{L}$ contains noncanonical kinetic terms but share the same field configuration $\phi(t)$, scale factor $a(t)$, energy density $\rho$, pressure $p$, and acceleration parameter $q$ with the canonical model.

\section{Twinlike models for the canonical model}
\label{sectionCosTwinFour}
In this section, we use the aforementioned first-order formalism to construct twinlike models for the canonical model. We explicitly show that for a canonical model there exist infinite noncanonical models which have the same background solution and properties than the canonical one. For simplicity, we only display three types of twinlike Lagrangians.
\subsection{Type-1 model}
The Lagrangian of the first type of twinlike model reads
\begin{eqnarray}
\mathcal{L}=\mathcal{L}(X,\phi^I),
\end{eqnarray} where $X=\delta_{IJ}X^{IJ}=\frac12\delta_{IJ}\dot{\phi}^I \dot{\phi}^J$. Using the first-order equation~\eqref{phiW}, we obtain the following \emph{on-shell condition}:
\begin{eqnarray}
\label{EqOnshell1}
X=\frac12\delta_{IJ}\frac{\partial W}{\partial \phi^I}\frac{\partial W}{\partial \phi^J}.
\end{eqnarray}
To become a twinlike model of $\bar{\mathcal{L}}$, the noncanonical model has to satisfy the following on-shell equations:
\begin{eqnarray}
\label{Eqsonshell}
\phi|=\bar{\phi},\quad
p|=\bar p,\quad
\rho|=\bar\rho,\quad
a|=\bar{a},\quad
q|=\bar{q}.\quad
\end{eqnarray}
The symbol $|$ here means taking the on-shell condition \eqref{EqOnshell1}. Obviously, $\phi|=\bar{\phi}$ is already satisfied, because both $\phi$ and $\bar{\phi}$ satisfy the on-shell condition \eqref{phiW}, or equivelently, Eq.~\eqref{EqOnshell1}.

Thus, we only need to check the other four equations, let us start with $p|=\bar p$. From Eqs.~\eqref{EqpL} and \eqref{EqbarL} we know that $p|=\bar p$ is equivalent to
\begin{eqnarray}
\mathcal{L}|&=&\delta_{IJ}\dot{\phi}^I\dot{\phi}^J/2-V(\{\phi^I\})\nonumber\\
&=&X-V.
\end{eqnarray}
Then, from Eqs. \eqref{EqrhoL} and \eqref{EqbarLX}, we know that $\rho|=\bar{\rho}$ equivalents to
\begin{eqnarray}
\mathcal{L}_{X^{IJ}}|=\delta_{IJ}.
\end{eqnarray}
Since $\mathcal{L}_{X^{IJ}}=\frac{\partial \mathcal{L}}{\partial X}\frac{\partial X}{\partial X^{IJ}}=\mathcal{L}_X\delta_{IJ}$, the on-shell equation for the energy density is simply
\begin{eqnarray}
\label{EqLX}
\mathcal{L}_{X}|=1.
\end{eqnarray}
From equations \eqref{EqH}-\eqref{EqdotH}, we know that once $\phi|=\bar{\phi}$, $\rho|=\bar\rho$ and $p|=\bar p$ are satisfied, the last two equations $a|=\bar a$ and $q|=\bar{q}$ will be satisfied automatically. Because both $a$ and $\bar a$ can be solved by introducing the first-order equation \eqref{HW} along with the constraint \eqref{Eqconstraint}. Finally, the deceleration parameter is defined only by the scale factor, so one would have $q|=\bar{q}$, if $a|=\bar a$.

Now, we are ready to construct the first type twinlike model for $\bar{\mathcal{L}}$. One of the simple Lagrangian that satisfies all the on-shell equations \eqref{Eqsonshell} is
\begin{eqnarray}
\label{Eqtwin1}
\mathcal{L}=X-V+\sum_{i=2}^{+\infty} U_i\left(X-\frac{\delta^{IJ}}2\frac{\partial W}{\partial \phi^I}\frac{\partial W}{\partial \phi^J}\right)^i,
\end{eqnarray}
where $U_i=U_i(G_{IJ},X^{IJ},\phi^K)$ are some arbitrary functions.
Obviously, under the on-shell equation \eqref{EqOnshell1}, $\mathcal{L}|=\bar{\mathcal{L}}$ and $\mathcal{L}_{X}|=1$, and therefore, $\mathcal{L}$ describes a twinlike model for $\bar{\mathcal{L}}$. One should keep in mind, however, that $\mathcal{L}$ and $\bar{\mathcal{L}}$ are essentially two different models. For example, $\mathcal{L}_{XX}|=2U_2(G_{IJ},X^{IJ},\phi^K)\neq\bar{\mathcal{L}}_{XX}=0$. Such difference appears as soon as the linear perturbations are considered (see the discussion of Refs.~\cite{AdamQueiruga2012,ZhongLiu2015}).
\subsection{Type-2 model}
Let us move to another type of Lagrangain
\begin{eqnarray}
\mathcal{L}=\mathcal{L}(\tilde{X},X,\phi^I),
\end{eqnarray}
where
\begin{eqnarray}
\tilde{X}\equiv\frac12G_{IJ}(\phi^K)\frac{\partial W}{\partial \phi^I}\frac{\partial W}{\partial \phi^J}.
\end{eqnarray}
In this case, the symbol $|$ would represent two on-shell conditions for $\phi^I$:
\begin{eqnarray}
X=\frac12\delta_{IJ}\frac{\partial W}{\partial \phi^I}\frac{\partial W}{\partial \phi^J},\quad
\tilde{X}=\frac12 G_{IJ}\frac{\partial W}{\partial \phi^I}\frac{\partial W}{\partial \phi^J}.
\end{eqnarray}
Obviously, the condition $p|=\bar p$ still requires $\mathcal{L}|=X-V$, but now the condition $\rho|=\bar\rho$ imposes one more equation
\begin{eqnarray}
\mathcal{L}_{\tilde{X}}|=0,
\end{eqnarray}
in addition to Eq.~\eqref{EqLX}. A Lagrangian that satisfies all the on-shell equations is
\begin{eqnarray}
\label{Eqtwin2}
\mathcal{L}&=&X-V+\sum_{i=2}^{+\infty} c_i\left(X-\frac{\delta^{IJ}}2\frac{\partial W}{\partial \phi^I}\frac{\partial W}{\partial \phi^J}\right)^i\nonumber\\
&+&\sum_{i=2}^{+\infty} d_i\left(\tilde{X}-\frac{G^{IJ}}2\frac{\partial W}{\partial \phi^I}\frac{\partial W}{\partial \phi^J}\right)^i,
\end{eqnarray}
where both $c_i$ and $d_i$ are arbitrary functions of $G^{IJ}$, $X_{IJ}$ and $\phi^I$. Once the superpotential $W$ is specified, one can explicitly reexpress $\mathcal{L}$ in terms of $\tilde{X}$, $X$ and $\phi^I$.
\subsection{Type-3 model}
Now, we present the last type of twinlike model, whose Lagrange reads
\begin{eqnarray}
\mathcal{L}=\mathcal{L}(X^{IJ},G_{IJ}(\phi^K),X,\phi^I).
\end{eqnarray}
The on-shell symbol $|$ in this case represents the following conditions:
\begin{eqnarray}
X=\frac12\delta_{IJ}\frac{\partial W}{\partial \phi^I}\frac{\partial W}{\partial \phi^J},\quad
X^{IJ}=\frac12 \frac{\partial W}{\partial \phi^I}\frac{\partial W}{\partial \phi^J}.
\end{eqnarray}
To be a twinlike model of $\bar{\mathcal{L}}$, the Lagrangian $\mathcal{L}$ must satisfy the following on-shell equations:
\begin{eqnarray}
\mathcal{L}|=X-V,\quad \mathcal{L}_{X}|=1,\quad \mathcal{L}_{X^{IJ}}|=0.
\end{eqnarray}
One of the possible Lagrangian is
\begin{eqnarray}
\label{Eqtwin3}
\mathcal{L}&=&X-V+\sum_{i=2}^{+\infty} c_i\left(X-\frac{\delta^{IJ}}2\frac{\partial W}{\partial \phi^I}\frac{\partial W}{\partial \phi^J}\right)^i\nonumber\\
&+&\sum_{i=2}^{+\infty} d_i\left(X^{IJ}-2\frac{\partial W}{\partial \phi^I}\frac{\partial W}{\partial \phi^J}\right)^i.
\end{eqnarray}
Once again, $c_i$ and $d_i$ are arbitrary functions of $G^{IJ}$, $X_{IJ}$ and $\phi^I$. Note that there might be some residual field indices $I,J,\cdots$ in the second line of Eq.~\eqref{Eqtwin3}. These residual indices can be contracted by constructing suitable coefficient $d_i(G^{IJ},X_{IJ},\phi^I)$ with appropriate indices.
\section{Explicit solutions}
\label{sectionCosTwinFive}
In this section, we explore how the first-order formalism can be used to reproduce some important cosmological inflation models that have been reported in literature.
\subsection{Single field}
We first consider the case with $n=1$, and
\begin{eqnarray}
\mathcal{L}=\frac12\dot{\phi}^2-V(\phi).
\end{eqnarray}
The first-order equations are
\begin{eqnarray}
\dot{\phi}&=&\frac{\partial W}{\partial \phi},\\
H&=&-W+\alpha k Z,
\end{eqnarray}
In case $k\neq 0$, $W$ and $Z$ should satisfy the following constraint:
\begin{eqnarray}
\label{EqCons2}
 &&2(W-\alpha k Z)Z_\phi
 =
 \left(W_\phi Z_{\phi\phi}
 +Z_\phi W_{\phi\phi}\right),
\end{eqnarray}
where $W_\phi\equiv \frac{\partial W}{\partial\phi}$, $W_{\phi\phi}\equiv \frac{\partial^2 W}{\partial\phi^2}$, and so on.
The scalar potential reads
\begin{equation}
V = \frac{3}{2}{(W - \alpha kZ)^2} + \frac{1}{2}{W_\phi }\left( {3\alpha k{Z_\phi } - {W_\phi }} \right).
\end{equation}
Consider, for example,
\begin{equation}
\label{EqWphi}
W=m\phi,\quad \frac{\partial W}{\partial\phi}=m,
\end{equation}
where $m$ is a constant, then the constraint equation \eqref{EqCons2} reduces to
\begin{equation}
2(m\phi  - \alpha kZ){Z_\phi } = m{Z_{\phi \phi }}.
\end{equation}
Obviously, for $k\neq 0$
\begin{equation}
Z = \frac{m}{{\alpha k}}\phi.
\end{equation}
Therefore, the scalar potential is
\begin{equation}
V = \frac{3}{2}{m^2}{\phi ^2} - \frac{1}{2}{m^2},
\end{equation}
for $k=0$, and
\begin{equation}
V = {m^2},
\end{equation}
for $k\neq0$. An inflation model of this kind of potential with $k=0$ was discussed in the textbook~\cite{LiddleLyth2000}, so we will not repeat the discussions here. But note that when $k\neq 0$, Eq.~\eqref{HW} reads $H=0$ and therefore predicts a unfavorable static Universe.

Now that we have the explicit expression of $W$, we can write down the Lagrangian of the twinlike models in terms of the field itself, and for type-1 model the result is
\begin{equation}
{\cal L} = X - V + \sum\limits_{i = 2}^{ + \infty } {{U_i}} {\left( {X - \frac{{{m^2}}}{2}} \right)^i},
\end{equation}
where $X=\frac12\dot{\phi}^2$ and $U_i$ are some arbitrary functions of $X$ and $\phi$. For the single field system here, one can show that the type-2 and type-3 twinlike models do not offer new result than the type-1 model.
\subsection{Multi-field models}
Using the first-order formalism in Sec.~\ref{sectionCosTwinThree} we can also reproduce some multi-field inflation models. For example, when $k=0$ the superpotential
\begin{eqnarray}
\label{EqWExp}
W\propto w_0 \exp(\lambda_I \phi^I)
\end{eqnarray}
would lead to the following scalar potential:
\begin{eqnarray}
V\propto  \exp(\lambda_I \phi^I).
\end{eqnarray}
Here $\lambda_I$ are $n$ constant coefficients. This potential describes the \emph{generalized assisted inflation} model~\cite{CopelandMazumdarNunes1999,KantiOlive1999} (see~\cite{LiddleMazumdarSchunck1998,MalikWands1999} for the original assisted inflation model). While, by taking
\begin{eqnarray}
W\propto e^{-\phi_1}f(\phi_2),
\end{eqnarray}
one would obtain the so-called \emph{soft inflation} model~\cite{BerkinMaedaYokoyama1990}.

Once the superpotential $W$ is specified, it is straightforward to use Eqs.~\eqref{Eqtwin1}, \eqref{Eqtwin2} and \eqref{Eqtwin3} to write down the Lagrangian of the corresponding twinlike models.

The construction of models with $k\neq0$ is rather difficult due to the complexity of the constraint equation \eqref{Eqconstraint}. But in principle, for some simple form of $W$, it is possible to find the solution of $Z$. After $W$ and $Z$ are obtained, one can immediately write the scalar potential $V$. A complete investigation on models with $n\geq 2$ and $k\neq 0$ is beyond the scope of the present work. We would leave it for our future work.

\section{Summary and discussions}
\label{sectionCosTwinSix}
The aim of this work is to construct twinlike models for cosmological models with $n$ canonical scalar fields. By using a new first-order formalism, we showed that it is possible to establish different kinds of twinlike models for a given canonical model with arbitrary integer $n$, regardless the spatial curvature is vanished or not. This compensate the work of Ref.~\cite{BazeiaDantas2012}, which failed to construct cosmological twinlike models for $k\neq 0$. In fact, it is the first-order formalism used by the authors of Ref.~\cite{BazeiaDantas2012} that prohibits the existence of the twinlike models for $k\neq0$.

We reproduced inflation models both for $n=1$ and $n>1$. For $n=1$ case, a linear superpotential $W\propto \phi$ can reproduce a quadratic scalar potential $V\propto \phi^2$ for $k=0$, and a constant potential for $k\neq 0$. The former describes a typical inflation model, while the later describes a disfavored static Universe. For $n>2$ case, we showed that when $k=0$, it is possible to reproduce the generalized assisted inflation and the soft inflation models by choosing particular superpotentials. The explicit Lagrangians for the corresponding twinlike models can be easily obtained by simply substituting the superpotentials into Eqs.~\eqref{Eqtwin1}, \eqref{Eqtwin2} and \eqref{Eqtwin3}.

We didn't offer explicit models in the case with $n>2$ and $k\neq0$, which deserves for a further consideration. In addition to inflation, the first-order formalism in the present work might also be useful for the study of other cosmological issues such as dark energy.
\section*{Acknowledgments}

This work was supported by the National Natural Science Foundation of China (Grants No. 11375075, No. 11522541 and No. 11405121).

%


\end{document}